\documentclass[letterpaper, 10 pt, conference]{ieeeconf}  

\IEEEoverridecommandlockouts                              

\usepackage{multirow}
\usepackage{amsmath}
\usepackage{subcaption}
\usepackage{xcolor}

\usepackage{graphics} 
\usepackage{epsfig} 
\usepackage{times} 
\usepackage{amsmath} 
\usepackage{amssymb}  
\usepackage{nicefrac}
\usepackage{cite}

\usepackage{scalerel}
\usepackage{graphicx}
\graphicspath{{./figures/}}
\usepackage{placeins}
\usepackage{xcolor}

\usepackage{euler}
\usepackage{newtxtext, newtxmath}
\usepackage{lipsum}

\newcommand{\norm}[1]{\ensuremath{\left\| #1\right\|}}

\usepackage{background}
\SetBgContents{Copyright (c) 2024 IEEE}
\SetBgScale{1}
\SetBgAngle{0}
\SetBgPosition{current page.north east}
\SetBgHshift{-2.5cm}
\SetBgVshift{-1cm}

\title{\LARGE \bf Time-Robust Path Planning with \\ Piece-Wise Linear Trajectory for Signal Temporal Logic Specifications}

\author{Nhan-Khanh Le$^{1}$, Erfaun Noorani$^{2}$, Sandra Hirche$^{1}$, and John S. Baras$^{2}$
\thanks{$^{1}$NK. Le and S. Hirche are with the Chair of Information-oriented Control (ITR), School of Computation, Information and Technology, Technical University of Munich, 80333 Munich, Germany. Emails: {\tt\small \{nhankhanh.le, hirche\}@tum.de}.}%
\thanks{$^{2}$E. Noorani and J. Baras are with the Department of Electrical and Computer Engineering and the Institute for Systems Research (ISR) at the University of Maryland, College Park, MD, USA. Emails: {\tt\small \{enoorani,baras\}@umd.edu}.}%
\thanks{Research partially supported by ONR grant N00014-17-1-2622, by a grant from the Army Research Lab, by the Clark Foundation, and the European Union’s Horizon Europe innovation action program under grant agreement No. 101093822, "SeaClear2.0”.}
}

\usepackage{svg}
\usepackage{tikz}
\usetikzlibrary{automata,positioning}
\usepackage{multirow}
\usepackage{multicol}

\let\labelindent\relax 
\usepackage{enumitem}
\let\implies \Rightarrow
\usepackage{algorithm}
\usepackage{algpseudocode}
\usepackage{mathtools}

\newtheorem{remark}{Remark}
\newtheorem{definition}{Definition}

\newtheorem{example}{Example}

\newtheorem{theorem}{Theorem}

\newtheorem{problem}{Problem}

\newcommand{\set}[1]{\mathcal{#1}}
\newcommand{\nSet}[1]{\mathbb{#1}}

\def \intersect{\cap}

\def \lgTrue{\top}
\def \lgFalse{\perp}
\def \lgIff{\Leftrightarrow} 
\def \lgEqual{\Coloneqq}

\def \lgSatisfy{\models}

\def \lgAnd{\wedge}
\def \lgOr{\vee}
\def \eta {\text{ETA}}
\def\sign{\text{sgn}}
\def\hequal{\equiv}

\def\lgSpec{\phi}
\def\lgPred{\pi}
\def\lgNeg{\lnot}
\def\tlNot{\lgNeg}
\def\tlAlways{\square}

\def\tlEvtll{\Diamond}

\def\tlUntil{\mathcal{U}}
\newcommand{\stlEvtll}[1]{\tlEvtll_{[#1]}}
\newcommand{\stlAlways}[1]{\tlAlways_{[#1]}}
\newcommand{\stlUntil}[1]{\tlUntil_{[#1]}}

\def\trob{\theta_{\bowtie}}
\newcommand{\trobzj}[1]{\theta_{\bowtie,{#1}}}
\newcommand{\rtrobzj}[1]{\theta_{+,{#1}}}

\def \bigO {\mathcal{O}}
\newcommand{\tuple}[1]{\left<#1\right>}
\def \polyAb {\mathcal{P}(A,h)}
\def \infmum {\sqcap}
\def \supmum {\sqcup}
\def \binfmum {\bigsqcap}
\def \bsupmum {\bigsqcup}

\newenvironment{nalign}[1][]{
    \begin{equation} 
    \begin{aligned} 
}{
    \end{aligned} 
    \end{equation} 
    \ignorespacesafterend
}

\def \defkpeq {\max \{l \in \nSet{N}_K | ~\sign(z_i^{\pi}) = \sign(z^{\pi}_{i^\prime}), ~\forall i^\prime, i+1 \leq i^\prime \leq i+l\}}
\def \defkmeq {\max \{l \in \nSet{N}_K | ~ \sign(z_i^{\pi}) = \sign(z^{\pi}_{i^\prime}), \forall i^\prime, i-l \leq i^\prime \leq i-1\}}

\usepackage[acronym,style=alttree, toc=true, shortcuts, xindy, nomain, nonumberlist]{glossaries}
\newacronym{VaR}{VaR}{Value at Risk}
\newacronym{CVaR}{CVaR}{Conditional Value at Risk}
\newacronym{MILP}{MILP}{Mixed-Integer Linear Program}
\newacronym{MIQP}{MIQP}{Mixed-Integer Quadratic Programming}
\newacronym{SQP}{SQP}{Sequential-Quadratic Programming}
\newacronym{CBF}{CBF}{Control Barrier Function}
\newacronym{GP}{GP}{Gaussian Process}
\newacronym{MRTA}{MRTA}{Multi-Robot Task Assingment}
\newacronym{MAS}{MAS}{Multi-Agent System}
\newacronym{SAS}{SAS}{Single-Agent System}
\newacronym{MAPF}{MAPF}{Multi-Agent Path Finding}
\newacronym{MRPF}{MRPF}{Multi-Robot Path Finding}
\newacronym{MDP}{MDP}{Markov Decision Process}
\newacronym{POMDP}{POMDP}{Partially-Observable Markov Decision Process}
\newacronym{PRM}{PRM}{Probabilistic Road Map}
\newacronym{RRT}{RRT}{Rapid-Exploring Random Tree}
\newacronym{PWL}{PWL}{Piece-Wise Linear}
\newacronym{MPC}{MPC}{Model Predictive Control}
\newacronym{PID}{PID}{Proportional – Integral – Derivative}
\newacronym{ETA}{ETA}{Estimated Time-of-Arrival}
\newacronym{GPS}{GPS}{Global Positioning System}
\newacronym{IMU}{IMU}{Inertial Measurement Unit}
\newacronym{LIDAR}{LIDAR}{Inertial Measurement Unit}

\newacronym{STL}{STL}{Signal Temporal Logic}
\newacronym{MTL}{MTL}{Metric Temporal Logic}
\newacronym{MITL}{MITL}{Metric Interval Temporal Logic}
\newacronym{LTL}{LTL}{Linear Temporal Logic}
\newacronym{TL}{TL}{Temporal Logic}

\newacronym{CPS}{CPS}{Cyber-Physical Systems}

\newacronym{UAV}{UAV}{Unmanned Aerial Vehicle}
\newacronym{UGV}{UGV}{Unmanned Ground Vehicle}
\newacronym{CBS}{CBS}{Conflict-Based Search}
\newacronym{GCS}{GCS}{Ground Control system}

\newacronym{MBSE}{MBSE}{Model-Based-Systems-Engineering}
\newacronym{SAR}{SAR}{Search-And-Rescue}
\newacronym{IB}{IB}{Instance-based}

\begin{document}

\maketitle
\thispagestyle{empty}
\pagestyle{empty}

\begin{abstract}
Real-world scenarios are characterized by timing uncertainties, e.g., delays, and disturbances. Algorithms with temporal robustness are crucial in guaranteeing the successful execution of tasks and missions in such scenarios. We study time-robust path planning for synthesizing robots' trajectories that adhere to spatial-temporal specifications expressed in Signal Temporal Logic (STL). In contrast to prior approaches that rely on {discretize}d trajectories with fixed time-steps, we leverage Piece-Wise Linear (PWL) signals for the synthesis. PWL signals represent a trajectory through a sequence of time-stamped waypoints. This allows us to encode the STL formula into a Mixed-Integer Linear Program (MILP) with fewer variables. This reduction is more pronounced for specifications with a long planning horizon. To that end, we define time-robustness for PWL signals. Subsequently, we propose quantitative semantics for PWL signals according to the recursive syntax of STL and prove their soundness. We then propose an encoding strategy to transform our semantics into a MILP. Our simulations showcase the soundness and the performance of our algorithm. 
\end{abstract}

\section{Introduction
}
{\ac{TL} is a branch of logic equipped with an expressive language capable of reasoning logical statements with respect to \emph{time} \cite{Baier2008}}. \ac{TL} is used to express the spatial-temporal behavior of a system unambiguously from natural languages, e.g., a mission specification ``robot $1$ is \emph{eventually} inside region $A$ and stays there for $5$ time units" can be captured unambiguously with \ac{TL} formulas.
Using \ac{TL} allows for synthesizing robots' trajectories with automated algorithms that meet the expected spatial-temporal behavior. This is known as the \emph{formal synthesis} problem. 

\ac{STL} is a variant of \ac{TL}. It extends its predecessor temporal logic languages, such as Linear Temporal Logic (LTL), to enable specifications of continuous-time signals. Its support for signals over the reals and its quantitative semantics \cite{Maler2004MonitoringTP} allow evaluating the \emph{robustness}, i.e., \emph{degree of satisfaction}, of a signal with respect to a specification. In other words, one can measure how well the system meets or falls short of meeting a specification. This is essential for practical automated mission design and synthesis to account for potential uncertainties. 

Exploring the robustness with respect to \ac{STL} formulas involves an assessment of both spatial and temporal dimensions. 
Formal \ac{STL} synthesis with spatial robustness is well-studied \cite{Raman2015ReactiveSF, Pang2017Smooth, Lindemann2019ControlBF}. Here, we focus on the notion of {temporal robustness}. 

Temporal robustness is essential in practical applications, where assigning fixed-time missions is challenging and robots frequently encounter delays, unexpected early starts, or  
disturbances during execution. The notion of temporal robustness for \ac{STL} formulas was introduced in \cite{Donze2010RSTL}. Following this, \cite{akazaki2015time, rodionova2021timerobust, Rodionova2022TRTL} further investigated the notion of temporal robustness in the context of monitoring and control synthesis. 

One approach employed for formal synthesis with \ac{STL} is to transform (space and time) robust \ac{STL} with linear predicates into linear constraints using mixed binary and continuous variables and construct a \ac{MILP}. In \cite{rodionova2021timerobust, Rodionova2022TRTL}, the authors propose a strategy to encode time robustness into linear constraints for {discretize}d signals, which allows for time-robust control synthesis using \ac{MILP}s. This builds upon prior methods in \cite{Karaman2008OptimalCO, Raman2015ReactiveSF} that propose the encoding rules using big-M methods \cite{griva2009linear} to encode space robust \ac{STL} with linear predicates into a \ac{MILP}. This approach does not scale well. The generated constraints from an \ac{STL} formula are imposed on \emph{each} {discretize}d trajectory instance. This implies that the \ac{MILP} increases in the number of binary and continuous variables proportionally to the planning horizon. Several methods for achieving an efficient optimization-based synthesis have been proposed, ranging from sequential quadratic programming \cite{Pang2017Smooth} and gradient-based \cite{Leung2020BackpropagationTS} methods to using time-varying \ac{CBF} and decomposition \cite{zhang2023modularized}.
%
%
%
%
%
However, these methods are investigated solely for space-robustness. Since the definition of time-robustness involves time-shifting, the above-mentioned methods are not applicable \cite{rodionova2021timerobust}. 

This motivates us to \textit{explore a scalable synthesis method that also provides robustness against timing uncertainties}. Our method is inspired by the hierarchical path-planning method using \ac{PWL} signals proposed in \cite{Chuchu2020FGS, Sun20223451}. A \ac{PWL} signal considers a trajectory as a sequence of waypoints consisting of timestamps and values. As a result, the size of the constructed \ac{MILP} does not depend on the time horizon but solely on the number of waypoints. \cite{Chuchu2020FGS, Sun20223451} show that various planning problems can be represented by a fewer number of waypoints compared to the number of {discretize}d trajectory instances, especially for long-time-horizon planning problems. The notion of time-robustness for \ac{PWL} signals has not been investigated. We propose and investigate the concept of time-robustness for \ac{PWL} signals. We summarize our contribution in detail here:

\noindent \textbf{Contributions \& Organization.} 
\begin{itemize} [leftmargin = *]
    \item \emph{Definition and Quantitative Semantics of Time-Robust \ac{STL} for PWL Signals} (Section \ref{sec:mr}): 
    We define the notion of temporal robustness with respect to an STL specification for \ac{PWL} signals. We then propose the quantitative semantics and prove their soundness. With our semantics, we are able to express time-robustness directly using the few numbers of waypoints, and evaluate the time-margin of a \ac{PWL} signal with respect to a specification.  
    \item \emph{Time-Robust \ac{STL} Path Planning} (Section \ref{sec:pf}): We formulate a hierarchical time-robust formal synthesis problem using \ac{PWL} signals. This allows for synthesizing the waypoints that not only satisfy a specification but also maximize the time-robustness.
    \item \emph{MILP Encoding} (Section \ref{sec:milp_enc}): We propose an encoding strategy to transform our proposed semantics into a \ac{MILP}. This allows for direct solutions to our formulated time-robust path planning problem.
    \item \emph{Simulation Results} (Section \ref{sec:bm}): We implement the synthesis algorithm and use several different mission specifications with different complexity to verify the soundness and study the performance of our method.
\end{itemize}

%
%

\textbf{Mathematical Notation}: True and False are denoted by $\lgTrue$ and $\lgFalse$, respectively. We use the set ${\nSet{B} = \{\lgFalse, \lgTrue\} = \{0, 1\}}$ to denote both sets of boolean and binary values. The set of natural, integer, and real numbers are denoted by $\nSet{N}, \nSet{Z}, \nSet{R}$, respectively. We use $\nSet{Z}_{\geq 0}$ and $\nSet{R}_{\geq 0}$ to denote the set of non-negative integers and real numbers. The set of strict positive natural index is denoted by ${\nSet{N}_K:=\{1, 2, \cdots, K\}}$. We use $\infmum$ for infimum and $\supmum$ supremum operators. The assignment operator is $\lgEqual$. The Euclidean norm is denoted by ${\norm{.}_2}$. The ``equivalence" of expressions is denoted with $\hequal$. Additionally, we define the sign operator $\sign(\cdot)$ over boolean variables, i.e., $\sign(x)= 1$ when $x=1$ and $\sign(x)=-1$ when $x=0$.
\section{Preliminaries} \label{sec:prel}
We begin by introducing the syntax and qualitative semantics of \ac{STL} for continuous signals. Given the inherent computational challenges in handling continuous signals using the existing solution methods, we opt for \ac{PWL} signals as a practical alternative. We present the definition of \ac{PWL} signals and their qualitative semantics.

\subsection{\ac{STL}}
We first present the syntax and semantics of \ac{STL} for continuous signals \cite{Donze2010RSTL}. Let $\xi(\cdot): \nSet{T} \xrightarrow{} \nSet{X}$ be a continuous signal where $\xi(t) \in \nSet{X} \subseteq \nSet{R}^d$  denotes the real-valued signal instance with $d$ dimensions at time instant $t \in \nSet{T} \subseteq \nSet{R}_{\geq 0}$. 
%
Within the context of motion planning for robots, we use the terms \emph{signal} and \emph{trajectory} interchangeably. 
\begin{definition}[STL Syntax]  \label{def:stl_syntax} 
\begin{align} \label{eq:stl_syntax}
    \lgSpec \lgEqual  {\lgPred} | \lgNeg {\lgPred} | \lgSpec_1 \lgAnd \lgSpec_2 | \lgSpec_1 \lgOr \lgSpec_2 | \tlEvtll_{I} \lgSpec | \tlAlways_{I} \lgSpec | \lgSpec_1\tlUntil_{I}\lgSpec_2,  
\end{align}
where ${\lgPred}$ and ${\lgNeg\lgPred}$ denote an atomic predicate and its negation,
and $\lgSpec_1$ and $\lgSpec_2$ are formulas of class $\lgSpec$. {The logical operators are conjunction ($\lgAnd$) and disjunction ($\lgOr$).} Temporal operators are ``Eventually" ($\tlEvtll$), ``Always" ($\tlAlways$) and ``Until" ($\tlUntil$) with the subscript $I$ denoting the time interval ${[a,b], 0 \leq a \leq b}$. 
\end{definition} 
Similar to \cite{Sun20223451}, we use \emph{Negation Normal Form} syntax, i.e., negation is applied only to atomic predicates ${\lgPred}$. It is not restrictive because we can express any \ac{STL} formula using this syntax \cite{Sun20223451}. We assume that an atomic predicate can be represented by a linear function. Let ${\mu(\cdot): \nSet{X} \xrightarrow{} \nSet{R}}$ be a linear function that maps a signal instance to a real value. Such function represents an atomic predicate ${\lgPred}$ through the relationship ${{\lgPred} \lgEqual \mu(\xi(t)) \geq 0}$, i.e., ${\lgPred}$ holds \emph{true} if and only if $\mu(\xi(t)) \geq 0$. 

In the following, we introduce the qualitative semantics of \ac{STL} which define when a signal satisfies a formula $\lgSpec$ at time $t$. In particular, we write $(\xi, t) \lgSatisfy \lgSpec$ if $\xi$ satisfies $\lgSpec$ at $t$, and 
$(\xi, t) \not\lgSatisfy \lgSpec$ if $\xi$ does not satisfy $\lgSpec$ at $t$. The semantics are defined \emph{recursively} for each formula in the syntax \eqref{eq:stl_syntax} below.

\begin{definition}[Qualitative Semantics of \ac{STL} formulas]
\begin{nalign} \label{eqn:qual_stl_semantics}
    (\xi, t) &\lgSatisfy {\lgPred} && \lgIff \mu(\xi(t)) \geq 0\\
    (\xi, t) &\lgSatisfy \tlNot{\lgPred} && \lgIff (\xi, t) \not\lgSatisfy {\lgPred}\\ 
    (\xi, t) &\lgSatisfy \lgSpec_1 \lgAnd \lgSpec_2 && \lgIff (\xi, t) \lgSatisfy \lgSpec_1 \lgAnd (\xi, t) \lgSatisfy \lgSpec_{2} \\ 
    (\xi, t) &\lgSatisfy \lgSpec_1 \lgOr \lgSpec_{2} && \lgIff (\xi, t) \lgSatisfy \lgSpec_1 \lgOr (\xi, t) \lgSatisfy \lgSpec_{2} \\ 
    (\xi, t) &\lgSatisfy \stlEvtll{a,b} \lgSpec && \lgIff \exists {{t^\prime} \in [t+a,t+b]} ~ (\xi, t^\prime) \lgSatisfy \lgSpec \\ 
    (\xi, t) &\lgSatisfy \stlAlways{a,b} \lgSpec && \lgIff \forall {{t^\prime} \in [t+a,t+b]} ~ (\xi, t^\prime) \lgSatisfy \lgSpec \\ 
    (\xi, t) &\lgSatisfy \lgSpec_1 \stlUntil{a,b} \lgSpec_2 && \lgIff \exists t^\prime \in [t + a, t + b], (\xi, t^\prime) \lgSatisfy \lgSpec_{2} \\ 
    & && \lgAnd 
    (\forall t^{\prime \prime} \in [t, t^\prime], (\xi,{t^{\prime\prime}}) \lgSatisfy\lgSpec_1) 
\end{nalign}  
\end{definition}
\subsection{\ac{PWL} Trajectory}
Rather than using a continuous signal to represent a robot's trajectory, one may use a sequence of timestamped waypoints with the assumption that the path between each pair of waypoints is linearly connected. We define \ac{PWL} signal below.
\begin{definition}[Waypoint]
A waypoint is a tuple consisting of a timestamp and the signal value  
\begin{align}
 w = \tuple{t, p}, t \in \nSet{T} \subseteq \nSet{R}_{\geq 0}, p \in \nSet{X} \subseteq \nSet{R}^d.   
\end{align}
\end{definition}
\begin{definition}[Linear Waypoint Segment]
A linear waypoint segment is a pair of a starting and an ending waypoint,
${z}_i = \{w_i, w_{i+1}\} \hequal \{\tuple{t_i, p_i}, \tuple{t_{i+1}, p_{i+1}}\}$ with ${0 \leq t_i \leq t_{i+1}}$. The value of the piece-wise linear signal at a given time $t$ can be represented by a linear function ${Z_i(\cdot): \nSet{T} \xrightarrow{} \nSet{X}}$, i.e.,
\begin{nalign}
Z_i(t) & = p_i + \frac{p_{i+1} - p_{i}}{t_{i+1} - t_i}(t - t_i),  t \in [t_i, t_{i+1}].
\end{nalign}
\end{definition}
\begin{definition}[\ac{PWL} Trajectory] \label{def:pwl_traj}
A piece-wise linear trajectory from $K$ waypoints is a sequence of $K-1$ linear waypoint segments
${\xi_{z_0} = \{z_{0}, z_{1}, \cdots, z_{K-2}\}}$. 
\end{definition}
\subsection{Qualitative \ac{STL} Semantics for \ac{PWL} Trajectory}
In line with prior research \cite{Sun20223451}, we are only interested in specifications that indicate the robots to reach and avoid particular regions with the expected temporal behavior. Therefore, we use atomic predicates ${\lgPred}$ and its negation ${\tlNot\lgPred}$ to particularly indicate whether a \emph{linear waypoint segment} lies \emph{inside} or \emph{outside} a region $\mathcal{P}$. We assume that a region $\mathcal{P}$ can be approximated by convex polygon ${\polyAb = \{p \in \nSet{R}^d ~| Ap \leq h\}, A \in \nSet{R}^{r \times d}, h \in \nSet{R}^r}$.

Let the binary variable $z^\lgSpec_i$ indicate the statement ``segment $z_i$ (or \emph{every} signal instance in $[t_i, t_{i+1}]$) satisfies the formula $\lgSpec$", i.e., ${z^\lgSpec_i \lgEqual z_i \lgSatisfy \lgSpec}$ and 
${z^\lgSpec_i = \lgTrue \lgIff \sign(z^\lgSpec_i) = 1 \lgIff (\xi, t) \lgSatisfy \lgSpec, \forall t \in [t_i, t_{i+1}]}$.
Then $\lgPred$ corresponds to region $\mathcal{P}$, i.e., $\lgPred$ holds true when the trajectory is ``inside $\mathcal{P}$" and $\lgNeg\lgPred$ when it is ``outside $\mathcal{P}$". 
From here on, we have $\pi$ and $\lgNeg\pi$ as the atomic predicates that indicate 
${z_i^\lgPred \lgEqual Z_i(t) \in \mathcal{P}, \forall t \in [t_i, t_{i+1}]}$, and ${z_i^{\lgNeg\lgPred} \lgEqual Z_i(t) \not\in \mathcal{P}, \forall t \in [t_i, t_{i+1}]}$. 

\begin{figure} 
    \centering
    \includegraphics[width = 0.6\linewidth]{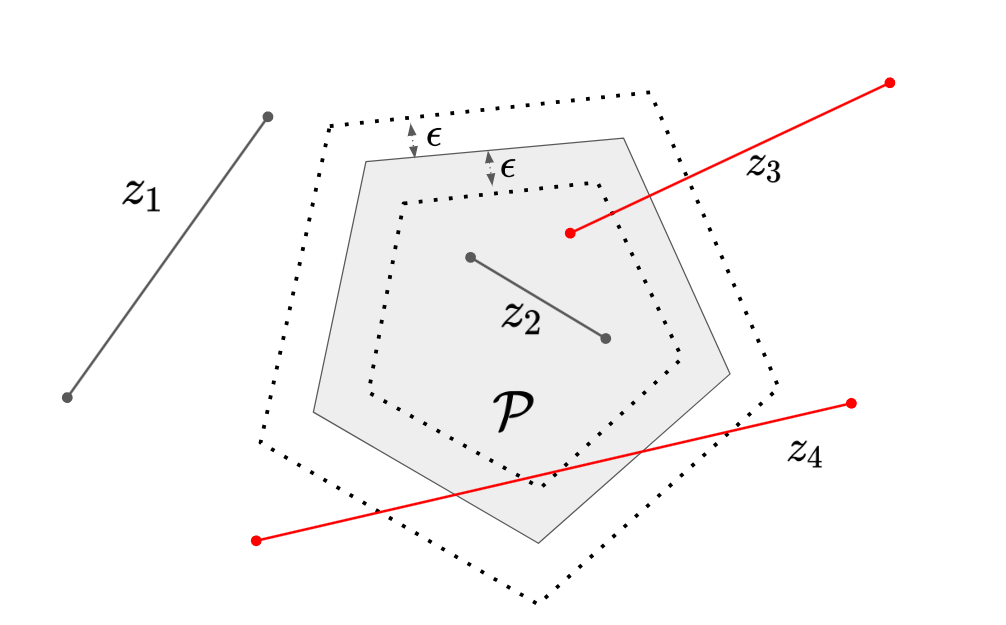}
    \caption{The gray area denotes the original polygon $\mathcal{P}$. The shrunk/bloated polygons by $\epsilon$ distant margin are the inner/outer dotted lines, respectively. The black line $z_2$ represents the line segment ``inside" of the shrunk region and the black line $z_1$ represents the line segment ``outside" of the bloated region. The red segments $z_3, z_4$ are neither ``inside" nor ``outside" $\mathcal{P}$.}  
    \label{fig:convex_poly}
\end{figure}

We now show that these ``atomic predicates" are constructed from conjunctions and disjunctions of the aforementioned linear predicate introduced in Definition \ref{def:stl_syntax}.
\begin{nalign} \label{eq:zpi_sdp}
     z_i^{{\lgPred}} \lgEqual &\bigwedge_{j = 1}^r \bigg( \frac{h^j - A^jp_i}{\norm{A^j}_2} \geq \epsilon  \lgAnd \frac{h^j - A^jp_{i+1}}{\norm{A^j}_2} \geq \epsilon \bigg), \\ 
     z_i^{\lgNeg {\lgPred}} \lgEqual & \bigvee_{j = 1}^r \bigg( \frac{A^j p_i - h^j}{\norm{A^j}_2} \geq \epsilon \lgAnd \frac{A^j p_{i+1} -  h^j}{\norm{A^j}_2} \geq \epsilon \bigg),
\end{nalign}
where $A^j, h^j$ are the coefficients of $j$-th rows of $A$ and $h$, i.e., one edge of the polygon. The positive space margin $\epsilon \in \nSet{R}$ is used to encounter the uncertain spatial deviation. Note that each term of \eqref{eq:zpi_sdp} is \emph{linear} in $p_i$ or $p_{i+1}$. This is consistent with our assumption of using linear predicates. The following example explains the notation.
\begin{example} See Figure \ref{fig:convex_poly}. ${z_i \lgSatisfy \pi \lgIff Z_i(t) \in \mathcal{P}, \forall t \in [t_i, t_{i+1}]}$, ${z_i \lgSatisfy \neg\pi \lgIff Z_i(t) \not\in \mathcal{P}, \forall t \in [t_i, t_{i+1}]}$. We have $z_1 \not\lgSatisfy \pi$, ${z_1 \lgSatisfy \neg\pi}$, ${z_2 \lgSatisfy \pi}$, ${z_2 \not \lgSatisfy \neg \pi}$, ${z_3 \not\lgSatisfy \pi}$, ${z_3 \not\lgSatisfy \neg\pi}$, ${z_4 \not\lgSatisfy \pi}$, and ${z_4\not\lgSatisfy \neg\pi}$.
\end{example}

For other operators, \cite{Sun20223451} proposes the following qualitative semantics:
\begin{nalign} \label{eqn:pwl_stl}
    z_{i}^{\stlAlways{a,b}\lgSpec} &\lgEqual \bigwedge_{j = 1}^{K} \left([t_j, t_{j+1}] \intersect [t_{i} + a, t_{i+1} + b] \neq \varnothing \implies z_j^\lgSpec \right). \\
    z_{i}^{\stlEvtll{a,b}\lgSpec} &\lgEqual (t_{i+1} - t_i \leq b - a) \\ &\lgAnd  \bigvee_{j = 1}^{K} \left([t_j, t_{j+1}] \intersect [t_{i+1} + a, t_{i} + b] \neq \varnothing \lgAnd z_j^\lgSpec \right). \\
    z_{i}^{\lgSpec_1\stlUntil{a,b}\lgSpec_2} &\lgEqual (t_{i+1} -t_i \leq b -a) \lgAnd \\ 
    &\bigvee_{j = 1}^{K} \bigg ([t_j, t_{j+1}] \intersect [t_{i+1} +a, t_i +b] \neq \varnothing \lgAnd z_j^{\lgSpec_2} \\
    & \lgAnd \bigwedge_{l = 0}^j [t_l, t_{l+1}] \intersect [t_{i}, t_{i+1} +b] \neq \varnothing \implies z_l^{\lgSpec_1} \bigg). 
    %
\end{nalign}
The time-intersection in the above semantics is to guarantee that \emph{every} instance of a segment $z_i$ satisfies the \ac{STL} formula $\lgSpec$. For example, consider a formula with the ``always" operator $z_{i}^{\stlAlways{a,b}\lgSpec}$, we require the signal at \emph{every} time-instances ${t_i \in [t_i, t_{i+1}]}$ satisfy $\stlAlways{a,b}\lgSpec$, which is equivalent to requiring the signal at \emph{every} time-instance between ${[t_{i} + a, t_{i} + b]}$ and ${[t_{i+1} + a, t_{i+1} + b]}$ (or they can be combined into ${[t_{i} + a, t_{i+1} + b]}$) to satisfy $\lgSpec$. A similar explanation is applied for the other operators. Note that the semantics are stricter than necessary compared to \eqref{eqn:qual_stl_semantics} \cite{Sun20223451}.
\section{Time-Robust \ac{STL} for PWL Signal} \label{sec:mr}
\def \tr{\text{time robustness}}
Here, we present the definition of time robustness and its quantitative \ac{STL} semantics for \ac{PWL} signals. The degree of  time-robustness, as defined in the work of Donze et al. \cite{Donze2010RSTL}, quantifies the duration $\theta$ during which a signal maintains its characteristic with respect to an \ac{STL} formula. This leads to two notions of time-robustness: right and left time robustness. These measure the time margin after and before a specified time instance. In other words, a signal has $\theta^\lgSpec_+$ right-time robustness if it satisfies $\lgSpec$ for \emph{any} preceding time instance between $[t, t + \theta^\lgSpec_+]$. This allows the synthesized signal to encounter \emph{early start} (left-shifted up to $ \theta^\lgSpec_+$). Meanwhile, left time-robustness is achieved if the signal satisfies $\lgSpec$ for \emph{any} succeeding time instance between $[t - \theta^\lgSpec_-, t]$. This offers robustness to encounter \emph{delay} (right-shifted up to $ \theta^\lgSpec_-$). Similarly, we denote the right/left time robustness of a linear segment $z_i^{\lgSpec}$ with respect to $\lgSpec$ by $\trobzj{i}^\lgSpec$ where ${\bowtie = \{+,-\}}$.

\begin{definition} [Time-Robust STL for PWL Signal] \label{def:pwl_tr}
For atomic predicates ${\lgPred}$, the right $\theta_{+,i}$ and left $\theta_{-,i}$ time robustness of a linear segment $z_i$ is defined as
\begin{nalign} \label{eq:pwl_prop_tr_strict}
    \theta_{+,i}^{{\lgPred}} &= \sign(z_i^{\lgPred}) \sum_{j = i+1}^{i + k^+}(t_{j+1} - t_j),\\
    \theta_{-,i}^{{\lgPred}} &= \sign(z_i^{\lgPred}) \sum_{j = i-k^-}^{i-1}(t_{j+1} - t_j),
\end{nalign}
where $k^+ = \defkpeq$, and $k^- = \defkmeq$. For $k = 0$, we set $\theta_{+,i}^{{\lgPred}} = \theta_{-,i}^{{\lgPred}} = 0$.
\end{definition}

For negation, we use a similar definition but replace $z_i^{\lgPred}$ with $z_i^{\neg \lgPred}$ and $\trobzj{i}^{{\lgPred}}$ with $\trobzj{i}^{{\lgNeg\lgPred}}$. Intuitively, \eqref{eq:pwl_prop_tr_strict} aggregates the time interval of the succeeding/preceding segments $z_{i^\prime}$ as long as both $z_i$ and $z_{i^\prime}$ either satisfy $\lgPred$ or both violate it, i.e., ${\sign(z_i^{\pi}) = \sign(z^{\pi}_{i^\prime})}$. 

For logical and temporal operators in the recursive syntax \eqref{eq:stl_syntax}, we propose the following semantics: 
\begin{align}
    \trobzj{i}^{\lgSpec_1 \lgAnd \lgSpec_2} &= \trobzj{i}^{\lgSpec_1} \infmum \trobzj{i}^{\lgSpec_2}; ~\trobzj{i}^{\lgSpec_1 \lgOr \lgSpec_2} = \trobzj{i}^{\lgSpec_1} \supmum \trobzj{i}^{\lgSpec_2}. && \label{eq:log_time_rob_strict}\\ 
    \trobzj{i}^{\stlAlways{a,b}\phi} &= \binfmum_{j \in \set{J}^\intersect} \trobzj{j}^\phi, \label{eq:alws_time_rob_strict} \\ 
    \set{J}^\intersect &= \left\{j \in \nSet{N}_K | [t_j, t_{j+1}] \intersect [t_i + a, t_{i+1} + b]  \neq \varnothing\right\}. \notag\\
    \trobzj{i}^{\stlEvtll{a,b}\phi} &= \bsupmum_{j \in \set{J}^\intersect} \trobzj{j}^\phi, \label{eq:evtll_time_rob_strict}\\ 
    \set{J}^\intersect &= \left\{j \in \nSet{N}_K | [t_j, t_{j+1}] \intersect [t_{i+1} + a, t_i + b] \neq \varnothing\right\}, \notag\\
t_{i+1} + a &\geq t_i + b. \notag\\ 
     \trobzj{i}^{\phi_1\stlUntil{a,b}\phi_2} &= \bsupmum_{j \in \set{J}^\intersect} \left( \trobzj{j}^{\phi_2} \infmum \binfmum_{l \in \set{L}_j^\intersect}\trobzj{l}^{\phi_1} \right), \label{eq:until_time_rob} \\
    \set{J}^\intersect &= \left\{j \in \nSet{N}_K | [t_j, t_{j+1}] \intersect [t_{i+1} +a, t_i +b] \neq \varnothing\right\}, \notag\\ 
    \set{L}^\intersect_j &= \left\{l \in \nSet{N}_j | [t_l, t_{l+1}] \intersect [t_{i}, t_{i+1} +b] \neq \varnothing\right\}, \notag\\
    t_{i+1} + a &\geq t_i + b. \notag
\end{align}

We provide an explanation for the ``always" operator ~\eqref{eq:alws_time_rob_strict} similarly to the qualitative semantics \eqref{eqn:pwl_stl}, while the other operators can be understood similarly. The {\tr} $\trobzj{i}^{\stlAlways{a,b}\phi}$ is the minimal (infimum operator) {\tr} $\trobzj{j}^{\phi}$ of \emph{any} segments $z_j$ that intersects with the time interval ${[t_i +a, t_{i+1} +b]}$.  

The \emph{Soundness} property of our proposed semantics is formally stated in the following theorem. If segment $z_i$ has positive time robustness with respect to $\lgSpec$, we know that $z_i$ satisfies $\lgSpec$. Inversely, negative time robustness implies $z_i$ does not satisfy $\lgSpec$. The equality case is inconclusive.
\begin{theorem}[Soundness] \label{theorem:soundness}
    \begin{nalign}
        \trobzj{i}^{\phi} &> 0 \implies z_i^\phi = \lgTrue &&\hequal z_i \lgSatisfy \lgSpec, \\ 
        \trobzj{i}^{\phi} &< 0 \implies z_i^\phi = \lgFalse &&\hequal z_i \not\lgSatisfy \lgSpec. 
    \end{nalign}
    \begin{proof}
        See Appendix \ref{apx:sec_proof}.
    \end{proof}
\end{theorem}
Note that there might exist a segment $z_i$ that fulfills the specification $\lgSpec$, but does not guarantee a positive time-robustness, i.e., ${z_i \lgSatisfy \lgSpec \not \implies \trobzj{i}^{\phi} > 0}$. Also, the definition of time-robustness for atomic predicates in \eqref{eq:pwl_prop_tr_strict} depends on the qualitative semantics \eqref{eq:zpi_sdp}, which is conservative. In particular, there might exist a \ac{PWL} segment $z_i$ that has both its waypoints outside of a convex polygon, thus staying outside of the region, but \emph{do not have to fulfill} $z_i^{\lgNeg {\lgPred}} = \lgTrue$, i.e., ${\exists z_i, Z_i(t) \not\in \mathcal{P}, z_i^{\lgNeg\lgPred} = \lgFalse}$. This is to show that the proposed semantics are sound but not complete.

\section{Problem Formulation} \label{sec:pf}
Our goal is to generate robot trajectories that satisfy a spatial-temporal specification expressed with \ac{STL} and stay satisfied under time uncertainties. The direct synthesis from general \ac{STL} formula is not real-time capable \cite{Pant2018FlyByLogic, Zhou_2015_OMP}. To mitigate this issue, we adopt a \emph{hierarchical} planning framework, which involves an offline high-level planner that synthesizes a reference \ac{PWL} signal and a low-level controller for tracking. This control framework is commonly utilized in various autonomous systems \cite{Chuchu2020FGS, Sun20223451}. 

With the recursive syntax \eqref{eq:stl_syntax}, we are able to express the expected temporal behavior from the time reference $t_0 = 0$, i.e., $z_o$. In other words, the mission $\lgSpec$ is characterized \emph{qualitatively} by $z_o^\lgSpec$, and its time robustness \emph{quantitatively} by $\trobzj{0}^\lgSpec$. We formally state our time-robust synthesis problem below.
\begin{problem}[Time-Robust \ac{STL} Synthesis] \label{prob:opti} Given an \ac{STL} specification $\lgSpec$, find a sequence of $K$ waypoints to construct a \ac{PWL} signal ${\xi_{z_0}}$ that satisfies $\lgSpec$, and maximize the trajectory's {\tr} $\trobzj{0}^\lgSpec$. 
\begin{align*}
    \max_{\xi_{z_0}} -J(\xi_{z_0}) &+ \lambda \trobzj{0}^\lgSpec\\
    \textbf{s.t. ~~~~} \trobzj{0}^\lgSpec &\geq \theta^* > 0\\      
     |p_{i+1} - p_{i}| &\leq v_b (t_{i+1} - t_{i}), \forall i \in \{0, \cdots, K-1\}
\end{align*}
where the strict positive $\theta^*$ is the {{\tr} threshold}. The objective function $J(\xi_{z_0})$ expresses the desired path properties, e.g., minimal path length, or the shortest make-span. The time robustness is maximized through the term $\trobzj{0}^\lgSpec$. The positive scalar $\lambda $$>$$0$ is used to trade-off the time-robustness and performance. As $\lambda$ increases, it gives greater significance to ensuring time-robustness. The final inequality ensures the feasibility of the synthesized signal for dynamical systems by constraining the velocity to remain bounded, where $v_b $$>$$ 0$.
\end{problem}

\begin{remark}[Feasibility] \label{rm:assumption}
We assume the existence of a specific \emph{number of waypoints} $K$ for which the synthesis problem is \emph{feasible}. Moreover, if the solution exists, Theorem \ref{theorem:soundness} guarantees that it satisfies $\lgSpec$ because $\trobzj{0}^\lgSpec$ is strictly positive. 
In practice, $K$ is determined \emph{heuristically}, i.e., either given, or we increase $K$ incrementally until a feasible solution is obtained.
\end{remark}
\begin{remark}[Advantage of using \ac{PWL} Signals] \label{rm:pwl_vs_ib}
\ac{PWL} signal represents a robot's trajectory through a fixed number of $K$ waypoints. The parameter $K$ is \emph{mission dependent}, e.g., the environments, the number of reach-avoid regions, the spatial pattern of robots' trajectory, etc. To our understanding, $K$ is determined heuristically as there is no generic method that can pre-define the number of necessary waypoints. However, \cite{Chuchu2020FGS, Sun20223451} show that various expressive planning tasks in practice can be represented by a small $K$ compared to the \emph{number of instances} of a \emph{discretized signal}. Moreover, an advantageous property of \ac{PWL} signal over the discretized one is that \emph{$K$ does not directly depend on the planning horizon}. In particular, imagine that we use a discretized signal for a mission with a given planning horizon. For a long horizon, the signal, due to the lower bound of the sampling rate, needs more instances to cover the whole mission. The results of previous works \cite{Zhou_2015_OMP, Pant2018FlyByLogic} suggest that the large number of instances causes the computational burden of Problem \ref{prob:opti}. On the other hand, it is not necessary for \ac{PWL} signal to use more waypoints for a longer planning horizon. As long as the signal has sufficient waypoints (see Remark \ref{rm:assumption}), we can directly assign the planning horizon to the timestamp of the last waypoint. Therefore, we say that the number of waypoints is \emph{invariant} to the planning horizon and this is the primary benefit of \ac{PWL} signals, especially for long-horizon problems. We provide a more comprehensive complexity analysis in the next section. 
\end{remark}

\section{\acrfull{MILP}} \label{sec:milp_enc}
Problem \ref{prob:opti} can not be solved analytically for a general $\lgSpec$ due to the logical operators. Therefore, we propose an encoding strategy to transform it into a \ac{MILP}. Considering the timestamps and the spatial locations as optimization variables, the encoding uses additional binary and continuous variables to construct \emph{linear} constraints that correspond to $\trobzj{0}^\lgSpec$. This approach is widespread; see \cite{Karaman2008OptimalCO, Raman2015ReactiveSF, Zhou_2015_OMP, Rodionova2022TRTL}. 

We introduce the general big-M methods that encode the necessary operators. A detailed explanation of the encoding can be found in \cite{Karaman2008OptimalCO, Raman2015ReactiveSF}. We adapt the ideas in \cite{Rodionova2022TRTL} to encode time-robustness for atomic predicates with \ac{PWL} signals. Our contribution is the encoding for our proposed semantics.

\subsection{General \ac{MILP} Encoding} \label{sec:general_MILP}
To transform the \ac{STL} formula into a \ac{MILP}, the encoding aims to convert \empty{each operator} occurred in the semantics \eqref{eqn:pwl_stl}, and \eqref{eq:log_time_rob_strict}-\eqref{eq:until_time_rob}. We briefly summarize the necessary encoding that is applied in the previous works \cite{Karaman2008OptimalCO, Raman2015ReactiveSF, Rodionova2022TRTL} as follows
\begin{itemize} [leftmargin=*]
    \item \emph{Linear predicate:} Each linear predicate is represented by a linear function $\mu(\cdot)$. We use linear predicates to indicate whether \emph{a point} lies in the inside/outside face of an edge of a convex polygon $\mathcal{P}$ (see \eqref{eq:zpi_sdp}). For instance, let 
    ${\mu(p_i) = \frac{h^j - A^jp_i}{\norm{A^j}_2} - \epsilon}$ be a linear function. If $\mu(p_i) \geq 0$, $p_i$ is on the side of the edge $(A^j, h^j)$ that lies within the polygon (\emph{inside}) $\mathcal{P}$ with $\epsilon$ space margin. We can capture $\mu(p_i) \geq 0$  through a binary variable $b^{\mu}$ as
    \begin{flalign} \label{eq:milp_lin_pred}
    -M(1 - b^{\mu}) &\leq {\mu}(p_i) \leq Mb^{\mu} - m_\epsilon,
    \end{flalign}
    where $M, m_\epsilon$ are sufficiently large and small positive numbers, respectively. If $b^{\mu} = 1$, ${\mu}(p_i) \geq 0$, else ${\mu}(p_i) < 0$. For the case ``\emph{outside} $(A^j, h^j)$ by $\epsilon$ space margin", we apply the same encoding method with ${\mu(p_i) = \frac{A^jp_i - h^j}{\norm{A^j}_2} - \epsilon}$. 

    \item \emph{Conjunction}: To encode $b = \bigwedge_{i = 1}^m b_i$, i.e., conjunction of $m$ binary variables, we impose the following constraints ${b \leq b_i, \forall i \in \nSet{N}_m}$ \text{ and } ${b \geq \sum_{i = 1}^m b_i - (m-1).}$

    \item \emph{Disjunction}: To encode $b = \bigvee_{i = 1}^m z_i$, we impose ${b \geq b_i, \forall i \in \nSet{N}_m}$ \text{ and } ${b \leq \sum_{i = 1}^m b_i.}$
    
    \item \emph{Infimum and Supremum}: The infimum of $m$ continuous variables, i.e., $\theta = \binfmum_{i = 1}^m \theta_i$, is encoded by using $m$ binary variables $\{b_1, \cdots, b_m\}$ and impose the following linear constraints for $\theta$ as
    \begin{subequations} \label{eq:inf_sup}  
    \begin{align}        
        & \theta_i - (1 - b_i)M \leq \theta \leq \theta_i, \forall i \in \nSet{N}_m, \label{eq:inf_sup-a} \\ 
        &\sum_{i = 1}^{m} b_i = 1. \label{eq:inf_sup-b}
        \end{align}
    \end{subequations}
Equation \eqref{eq:inf_sup-a} ensures that $\theta$ is smaller than other variables and equal to any $\theta_i$ if and only if $b_i = 1$. Equation \eqref{eq:inf_sup-b} ensures that the infimum of the set exists. For supremum, we use similar encoding constraints as the infimum, but replace \eqref{eq:inf_sup-a} with ${\theta_i \leq \theta \leq \theta_i + (1 - b_i)M}$.

\item \emph{Product of continuous and binary variable}: To encode  ${y = xb, x \in \nSet{R}, b \in \nSet{B}}$, we impose
\begin{nalign} \label{eq:multi_xb}
  -Mb \leq y \leq Mb,~ x - M(1 - b) \leq y \leq x + M(1 - b),  
\end{nalign}
where $M>0$ is a sufficiently large upper bound of $x$.
\end{itemize}

\subsection{{Time Robustness} Encoding for \ac{PWL} signal} \label{sec:tre}
We use the right-time robustness to explain the encoding. A similar strategy can be applied for the left time robustness. We first introduce the time robustness encoding for the atomic predicates, and then for the temporal operators.
\begin{itemize} [leftmargin=*]
 \item \emph{Time robustness encoding for \eqref{eq:pwl_prop_tr_strict}, i.e., for $\lgPred$ and $\lgNeg \lgPred$}: Let $\rtrobzj{i}^{{\lgPred}}$ be the right {\tr} of segment $z_i$ with respect to an atomic predicate $\lgPred$, we adapt the ``counting" encoding concept proposed by \cite{Rodionova2022TRTL} to generate linear constraints such that $\rtrobzj{i}^{{\lgPred}}$ corresponds to the right {\tr} in \eqref{eq:pwl_prop_tr_strict}. To this end, we introduce two \emph{continuous} \emph{time-aggregating} variables $\Delta t^1_i$ and $\Delta t^0_i$  for \emph{each} segment and encode $\rtrobzj{i}^{{\lgPred}}$
\begin{align}
    \Delta t^1_i &= (\Delta t^1_{i+1} + \delta t_i)z_i^{\lgPred},
    \Delta t^0_i = (\Delta t^0_{i+1} - \delta t_i)(1 - z_i^{\lgPred})  \label{eq:t_aggre}\\
    \rtrobzj{i}^{{\lgPred}} &= \Delta t^1_i  + \Delta t^0_i + \left(- z_i^{\lgPred}\delta_i + (1 - z_i^{\lgPred})\delta_i \right) \label{eq:tr_raw}\\ 
    &= z_i^{\lgPred}\Delta t^1_{i+1} + (1 - z_i^{\lgPred})\Delta t^0_{i+1} \label{eq:right_time_rob},
\end{align}
with ${\delta t_i = t_{i+1} - t_i}$, ${\Delta t^1_{K-1} = \Delta t^0_{K-1} = 0}$.

According to \eqref{eq:pwl_prop_tr_strict}, the constraints depend on $z_i^\lgPred$, i.e., the qualitative value of segment $z_i$ with respect to the atomic predicate $\lgPred$. As defined in \eqref{eq:zpi_sdp}, $z_i^\lgPred$ indicates whether \emph{a segment} $z_i$ lies inside $\mathcal{P}$, which is constructed through conjunction and disjunction of linear predicates. Therefore, we apply the encoding techniques in Section \ref{sec:general_MILP} to obtain the constraints for $z_i^\lgPred$.

Next, we use $\Delta t^1_i$ and $\Delta t^0_i$ to encode the sign operator in \eqref{eq:pwl_prop_tr_strict}. In particular, \eqref{eq:t_aggre} implies that $\Delta t^1_i$ aggregates the time interval of the successive segments $z_{i+1}$ if ${z^\lgPred_{i+1} = z^\lgPred_{i} = 1}$, while  $\Delta t^0_i$ aggregates the negative time interval, i.e.,  $-\delta t_i$ if $\sign(z^\lgPred_{i+1}) = \sign(z^\lgPred_{i}) = 0$. Moreover, since \eqref{eq:t_aggre} is recursive, the aggregation continues until $\sign(z_{i+1}) \neq \sign(z_{i})$. This corresponds to the $\sum$ operator used in \eqref{eq:pwl_prop_tr_strict}. 

Finally, since these variables are disjunctive, i.e., ${\Delta t^1_i \neq 0 \implies \Delta t^0_i = 0}$ and vice versa \cite{Rodionova2022TRTL}, they can be combined into $\Delta t^1_i  + \Delta t^0_i$ as shown in \eqref{eq:tr_raw}. The last summand in \eqref{eq:tr_raw} is explained by the fact that the {\tr} defined in \eqref{eq:pwl_prop_tr_strict} does not contain the time interval $\delta_i$ of the current segment $z_i$, so we have to subtract or add $\delta_i$ according to the sign of $z_i$. The encoding for negation $\lgNeg \lgPred$ is performed similarly by replacing $z^\lgPred_{i}$ and $\rtrobzj{i}^{{\lgPred}}$  with $z^{\lgNeg\lgPred}_{i}$ and $\rtrobzj{i}^{{\lgNeg\lgPred}}$. 

Furthermore, we emphasize that \eqref{eq:t_aggre} and  \eqref{eq:right_time_rob} can be realized with linear constraints by using the encoding for multiplication of binary and continuous variables in \eqref{eq:multi_xb}.

\item \emph{Time robustness encoding for temporal operators \eqref{eq:log_time_rob_strict}-\eqref{eq:until_time_rob}}: We instruct the encoding for the ``eventually" ($\tlEvtll$), the other operators can be done similarly. Recall the expression from \eqref{eq:evtll_time_rob_strict}. Let $\trob^*$ be $\trobzj{i}^{\stlEvtll{a,b}\lgSpec}$. We introduce $K$$-$$1$ additional binary variables $\{b_1, \cdots, b_{K-1}\}$ and impose the following constraints
\begin{subequations} 
\begin{flalign}
& b_{ij}^\intersect \lgEqual (t_{i}+b - t_j \geq \epsilon_t) \lgAnd (t_{j+1} - (t_{i+1} +a) \geq \epsilon_t) \notag \\
& ~~~~~~ \lgAnd \left(t_{i+1} - t_i \leq b - a\right), \forall j \in \nSet{N}_{K-1}; \label{eq:timeintrs_evtll_tr}\\
& \trobzj{j}^\lgSpec - (1- b_{ij}^\intersect)M \leq \trob^*, \forall j \in \nSet{N}_{K-1}; \label{eq:geq_evtll_tr} \\
& \trob^* \leq \trobzj{j}^\lgSpec + (1- b_{j})M, \forall j \in \nSet{N}_{K-1}; \label{eq:evtll_tr_leq}\\  
& b_{ij}^\intersect \geq b_j, \forall j \in \nSet{N}_{K-1}; \sum_{j = 1}^K b_j = 1. \label{eq:supremum_exists}
\end{flalign}
\end{subequations} 
First, in \eqref{eq:timeintrs_evtll_tr}, the binary variable $b^\intersect_{ij}$ with $\epsilon_t \in \nSet{R}_{\geq 0}$ indicates whether two time-intervals $[t_j, t_{j+1}]$ and ${[t_{i+1} +a, t_{i} +b]}$ intersects. If $b^\intersect_{ij} = \lgTrue$, they intersect, else they don't. Equation \eqref{eq:geq_evtll_tr} implies that $\trob^*$ is greater or equal to the time robustness of \emph{any} segments that intersect with $[t_{i+1} +a, t_i + b]$. Meanwhile, \eqref{eq:evtll_tr_leq} guarantees that $\trob^*$ is bounded by $\trobzj{j}^\lgSpec$ whenever $b_{j} = 1$. Finally, \eqref{eq:supremum_exists} ensures that the supremum belongs to the set that intersects the interval, i.e., $b_j = 1 \implies b_{ij}^\intersect = 1$. Besides, $\sum_{j = 1}^K b_j = 1$ ensures that the supremum exists.
\end{itemize}

\subsection{Complexity Analysis} \label{sec:complex_ana}
We analyze the encoding complexity for a \ac{PWL} signal $\xi_{z_0}$ with $K$ waypoints.
Atomic predicates (AP) (or negations) are constructed through the set of $|AP|$ linear predicates, thus we need $\bigO(|AP|K)$ binary variables for $\xi_{z_0}$. Time robustness for AP requires $\bigO(1)$ additional continuous variables for each segment as shown in \eqref{eq:t_aggre}, thus $\bigO(K)$ for $\xi_{z_0}$. For temporal operators, we need additional variables to encode the time-intersection and infimum/supremum. Each segment requires $\bigO(K)$ additional binary and $\bigO(1)$ continuous variables, which respectively results in $\bigO(K^2)$ and $\bigO(K)$ for $\xi_{z_0}$. The generalized number of binary and continuous variables of the \ac{MILP} is $\bigO(|\lgSpec|K^2)$ and $\bigO(|\lgSpec|K)$, respectively \footnote{For a formula with \emph{one single temporal operator}, e.g., $\stlEvtll{I} {\lgPred}$, $\stlAlways{I} {\lgPred}$, we just impose constraints on the first segment, thus only requiring $\bigO(|AP|K)$ binary variables instead of $\bigO(|AP|K + |\lgSpec|K^2)$.}. 

It's worth highlighting a comparison between our results and the encoding achieved through discretizing the signal into $T$ instances. The analysis in \cite{Rodionova2022TRTL} shows a complexity of {$\bigO(|AP|T)$} binary and $\bigO(|\lgSpec|T)$ continuous variables. Although the number of binary variables required by \ac{PWL} signal depends on the quadratic term $K^2$, various path-planning problems can be represented in practice with $K^2 \ll T$ \cite{Sun20223451}. 
The analysis confirms Remark \ref{rm:pwl_vs_ib} that the encoding with \ac{PWL} signal \emph{does not directly depend on the planning horizon}. 



\subsection{Algorithm Overview}
We summarize our approach in the following algorithms. We initialize the \ac{PWL} signal ${\xi_{z_0}}$ as optimization variables. The \ac{STL} encoding is performed modularized for each region $\mathcal{P}$ and recursive for temporal operators (Algorithm \ref{alg:sr_ra}). Then the algorithm \ref{alg:all_in} combines the linear constraints obtained from the encoding of \emph{every} regions and additionally considers dynamical constraints to construct a complete \ac{MILP}. The generated \ac{PWL} trajectory/waypoints from an \ac{MILP} optimizer is then used as a reference for a low-level tracking controller.  

\def \cstrSetFinal{$\mathcal{X}_\lgSpec$}
\def \cstrSet{$\mathcal{X}$}
\begin{algorithm} 
\caption{Single Region Reach-Avoid Specification}
\label{alg:sr_ra}
\begin{algorithmic}[1]
\Require{Region $\mathcal{P}$, \ac{STL} Formula $\lgSpec$ for $\mathcal{P}$, number of waypoints $K$, variables of \ac{PWL} Trajectory ${\xi_{z_0}}$ 
from Definition \ref{def:pwl_traj}} 
\Ensure{Set of linear constraints \cstrSetFinal}
\State \cstrSetFinal $\gets$ \{\}
\State $A \in \nSet{R}^{m \times 2} , h \in \nSet{R}^{m}$ $\gets$ Convex Approximation of $\mathcal{P}$ 
\For{$j \gets 1$ to $m$}                    
    \For{$z_i^\lgPred \gets z_0^\lgPred$ to $z_{K-1}^\lgPred$}                    
        \State $z_i^\lgPred$, \cstrSet $\gets$ \eqref{eq:zpi_sdp}, $A^j, h^j$
        \State \cstrSetFinal $\gets$ \{\cstrSetFinal, \cstrSet\} 
    \EndFor 
\EndFor

\For{$z_i^\lgPred \gets z_0^\lgPred$ to $z_{K-1}^\lgPred$}   
    \State $\trobzj{i}^\lgPred$, \cstrSet  $\gets$ $z_i^\lgPred$, \eqref{eq:right_time_rob}
    \State \cstrSetFinal $\gets$ \{\cstrSetFinal, \cstrSet\}
\EndFor

\For{temporal operators in $\lgSpec$ (backward, recursive)}      \State $\trobzj{i}^\lgSpec$, \cstrSet $\gets$ Applies Section \ref{sec:milp_enc} for $\trobzj{i}^\lgPred$
    \State \cstrSetFinal $\gets$ \{\cstrSetFinal, \cstrSet\}
\EndFor
\end{algorithmic}
\end{algorithm}

\begin{algorithm} 
\caption{Time-Robust Synthesis with \ac{MILP}}
\label{alg:all_in}
\begin{algorithmic}[1]
\Require{STL Formula $\lgSpec$, number of waypoints $K$, maximal velocity $v_b > 0$, makespan $T > 0$} 
\Ensure{${\xi_{z_0}}$}
    \State Initialize decision variables ${\xi_{z_0}}$
    \State \cstrSetFinal $\gets$ \{\}
    
    \Statex{// \ac{STL} constraints}
    \For{ each region $\mathcal{P}$ in $\lgSpec$}   
    \State \cstrSet $\gets$ Algorithm \ref{alg:sr_ra}($\mathcal{P}$, $\lgSpec$, $K$, $\xi_{z_0}$) 
    \State \cstrSetFinal $\gets$ \{\cstrSetFinal, \cstrSet\}
    \EndFor

    \Statex{// Dynamic constraints for the waypoints}
    \State \cstrSetFinal $\gets$ \{\cstrSetFinal, $t_0 = 0, t_{K-1} = T$\}
    \Statex    
    {// Loop over each pair of waypoints $\tuple{t_i, p_i, t_{i+1}, p_{i+1}}$}
    \For{i $\gets$ 0 to $K-1$}   
    \State \cstrSetFinal $\gets$ \{\cstrSetFinal, $t_{i+1} \geq t_{i}, |p_{i+1} - p_{i}| \leq v_b (t_{i+1} - t_{i})$\}
    \EndFor
    \State{// \ac{MILP} Synthesis}
    \State ${\xi_{z_0}}$ $\gets$ GUROBI(\cstrSetFinal, $J({\xi_{z_0}})$)    
\end{algorithmic}
\end{algorithm}


\section{Benchmark} \label{sec:bm}
\begin{figure}[t]
\begin{multicols}{2}
    \subfloat[$\lgSpec_1$ Multi-Blocks-1
      \label{fig:exp-a}]{\includegraphics[width = 0.24\textwidth]{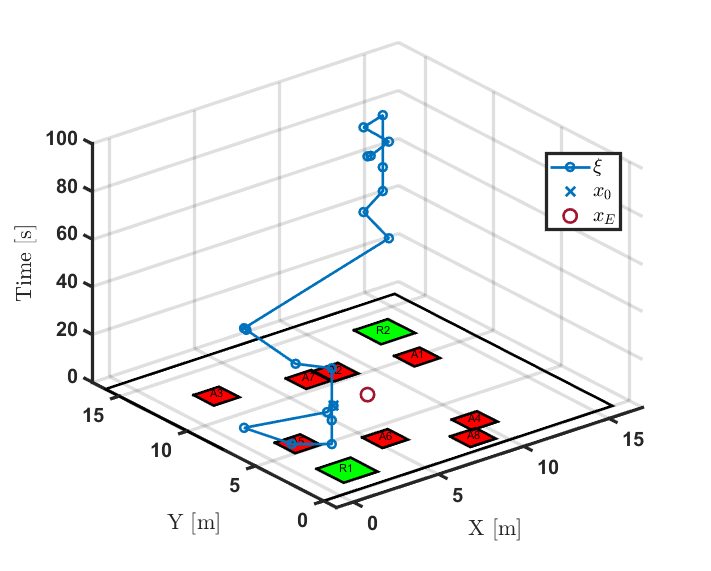}} \hfill \par
    \subfloat[$\lgSpec_2$ Multi-Blocks-2
      \label{fig:exp-b}]{\includegraphics[width = 0.24\textwidth]{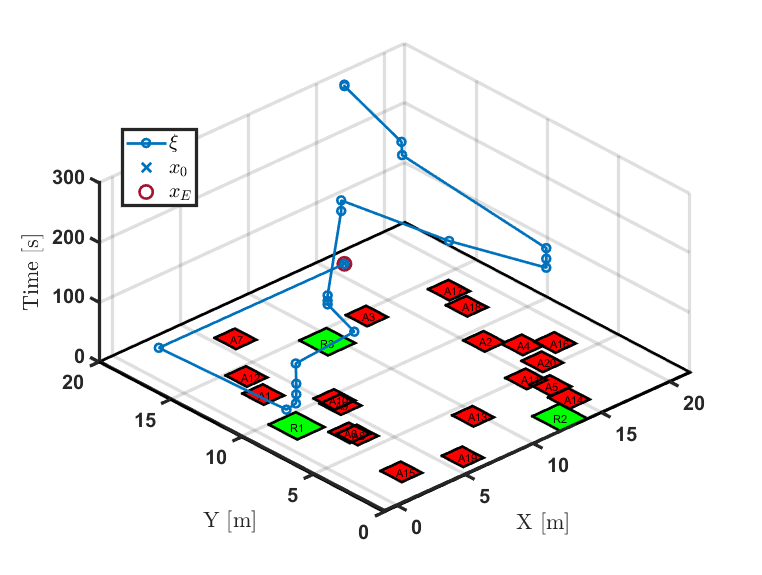}} \hfill \par
    \subfloat[$\lgSpec_3$ Door-Puzzle-1
      \label{fig:exp-dp1}]{\includegraphics[width = 0.24\textwidth]{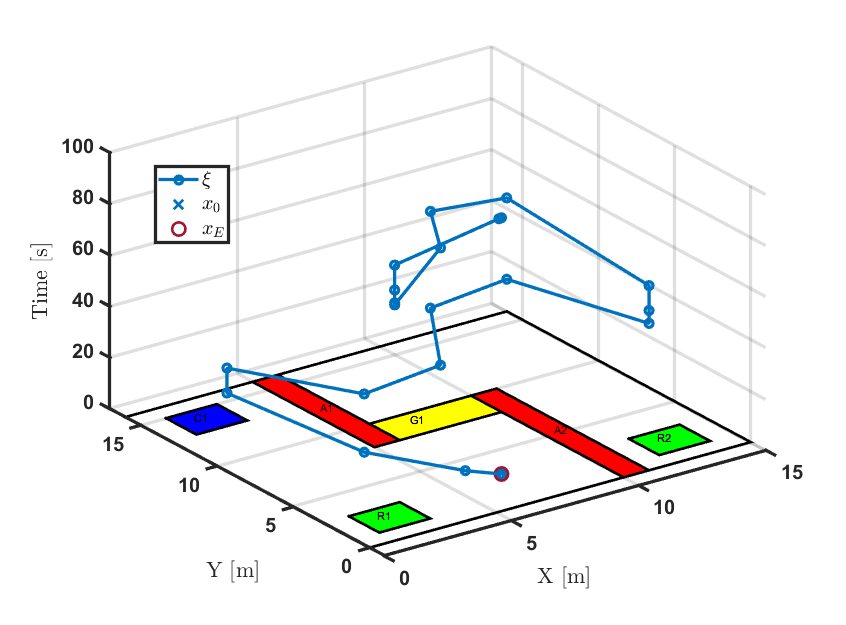}} \hfill \par
    \subfloat[$\lgSpec_4$ Door-Puzzle-2
      \label{fig:exp-dp2}]{\includegraphics[width = 0.24\textwidth]{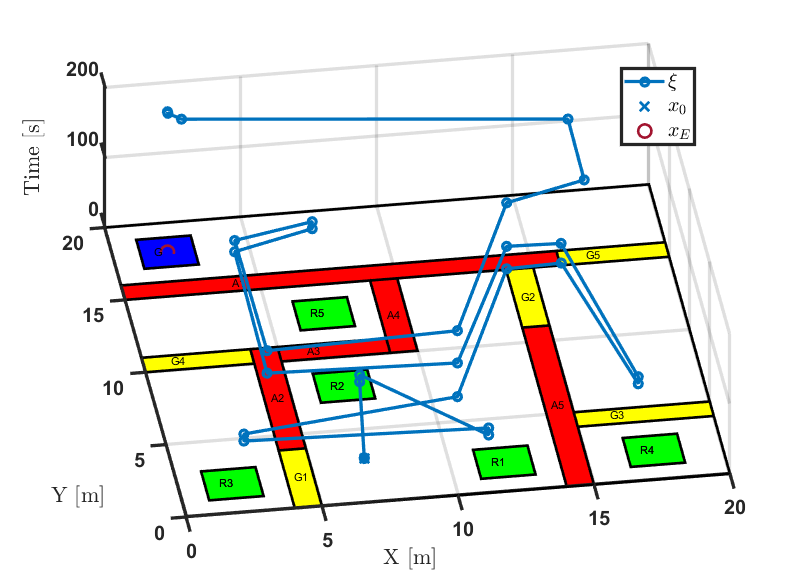}} \hfill \par
\end{multicols}
\caption{The simulation results of the four scenarios. The blue circles in the figures are the synthesized waypoints. The \ac{PWL} signal is the blue piece-wise linear path connecting the waypoints. The $XY$ plane consists of the reach (green/dark blue)-avoid (red/yellow) regions. The $Z$ axis denotes the synthesized time-stamp in seconds. The start and end points are marked as indicated in the legend.}
\label{fig:num_res}
\end{figure}

\subsection{Setup} To evaluate our method, %
we benchmark against the method using discretized signals (\ac{IB} method) in \cite{rodionova2021timerobust, Rodionova2022TRTL}
. To our knowledge, these are the sole studies in the synthesis of time-robust \ac{STL}. We implement both methods in the same hierarchical path-planning fashion, i.e., we use the encoding of \cite{rodionova2021timerobust, Rodionova2022TRTL} to synthesize the sequence of discretized robots' trajectory based on a single-integrator model with velocity constraint. 
We opted for a discretization step of $1$ s for the sake of argument simplicity. This choice aligns the planning horizon $T$ with the number of instances of the discretized signal. The two algorithms are tested with the same task-dependant planning time horizon and limited velocity of 1 m/s. For \ac{PWL} signal, we increase the number of waypoints $K$ incrementally until a feasible solution is obtained, and further increase $K$ to evaluate the potential increase in the objective function as a result of using more waypoints.

We evaluate our method in four example mission scenarios. The missions are chosen to showcase our algorithm in scenarios with a range of spatial and temporal complexity. 
{In all the missions, the objective is minimizing the path length encoded through $-J(\xi_{z_0})$ and maximizing the time-robustness $\trobzj{0}^\lgSpec$ with a trade-off parameter $\lambda = 1$} as shown in \ref{prob:opti}.
\subsection{Mission Scenarios}
Here, we explain the missions in detail first in plain English and then in \ac{STL} language. To simplify the notation, temporal operators without time subscripts imply the overall time horizon $[0, T]$, e.g., ${\tlEvtll \hequal \stlEvtll{0, T}}$.

%
In the first mission ``Multi-Blocks-1", the robot needs to visit a region and stay there for a certain time while avoiding the red regions all the time, i.e., ${\lgSpec_1 = \bigwedge_{i = 1}^{8} \lgNeg A_i \lgAnd  \tlEvtll \stlAlways{0,5} R_1  
\lgAnd \tlEvtll\stlAlways{0, 5} R_2}$. The starting and ending positions are given. 
In the second mission ``Multi-Blocks-2" with a longer mission horizon $T = 300 ~s$, the robot must monitor an additional region within a given time window, i.e., $\lgSpec_2 = \bigwedge_{i = 1}^{20} \lgNeg A_i \lgAnd  \stlEvtll{0, 150}\stlAlways{0,15} R_1  \lgAnd \stlEvtll{180, 260}\stlAlways{0, 15} R_2  
\lgAnd \stlEvtll{50,80}\stlAlways{0, 15} R_3 \lgAnd \stlAlways{200, 220} R_3$.

The third mission ``Door-Puzzle-1" requires the robot to eventually visit region $R_1$ and stay there for a certain time. Region $R_2$ must be eventually visited, but the robot is not allowed to pass through the yellow gate $G_1$ before reaching the charging area $C_1$, i.e., ${{\lgSpec_3 = \neg G_1\stlUntil{0, 30} C_1 \lgAnd \stlEvtll{50,80}\stlAlways{0,5}R_1 \lgAnd \tlEvtll\stlAlways{0,5}R_2}}$.
%

The fourth mission ``Door-Puzzle-2" is an example taken from \cite{Sun20223451}. The robot is not allowed to pass through the gates until the key with a corresponding number is collected at the green region. The final goal is to reach the blue region, i.e., 
${\lgSpec_4 = \tlEvtll G \lgAnd (\lgAnd_{i=1}^5 \tlAlways \lgNeg A_i) \lgAnd \lgAnd_{i=1}^5 \lgNeg D_i \tlUntil R_i}$.

\subsection{Simulation Results}
Table \ref{tab:bm} shows the encoded \ac{MILP}'s information and the optimization results. 
The missions are synthesized with various numbers of waypoints $K$, and the discretized method (the row corresponding to the planning horizon $T$). We compare the results \emph{row-wise} between the \ac{PWL} method with a fixed $K$ and the discretized method. 
Figure \ref{fig:num_res} depicts the synthesized waypoints and the linear segments. 
\begin{enumerate} [leftmargin=*]
    \item \emph{Soundness}: Overall, all obtained solutions are sound, i.e., none of them violate the imposed specification.

    \item \emph{Model Complexity}:
    \begin{itemize} [leftmargin=*]
        \item Encoding with \ac{PWL} signal requires fewer continuous variables (all four scenarios).
        \item \ac{PWL} signal requires more binary variables 
        as observed in $\lgSpec_1$ and $\lgSpec_3$. 
        For long planning horizons, e.g., $\lgSpec_2$ and $\lgSpec_4$, \ac{PWL} encoding requires fewer binary variables. 
    \end{itemize} 
    This is explained through the complexity analysis in Section \ref{sec:complex_ana}. The number of binary variables increases proportionally with $K^2$ ($\bigO (|\lgSpec|K^2)$), thus the encoding requires more binary variables when $T \ll K^2$. However, in $\lgSpec_2$, \ac{PWL} encoding uses fewer binary variables because these specifications are dominated by a large number of ``avoid" sub-formula with only one single temporal operator whose complexity is linear with $K$ ($\bigO (|AP|K)$) as explained in the footnote of Section \ref{sec:complex_ana}. 
    \item \emph{Objectives and Sensitivity Analysis}: 
    The objectives obtained from \ac{PWL} and \ac{IB} methods are approximately equivalent (in all scenarios $\lgSpec_1, \lgSpec_2$, and $\lgSpec_3$). The result of $\lgSpec_3$ shows that the objective depends on the number of waypoints $K$. First, we need at least $K $$=$$ 23$ to obtain a feasible solution with a poor time-robustness of 0.2 s. When we increase $K $$=$$ 25$, we achieve a better time-robustness (9.7 s). Nevertheless, further increasing $K$ \emph{does not} always improve the objective, which is observed when we further increase $K $$=$$ 30$, or when we vary $K$ in the experiments $\lgSpec_1$ and $\lgSpec_2$.
    This behavior can be explained by the fact that a greater number of waypoints expands the solution space and may contain a better solution. 
    \item \emph{Computation Time}: The solving time is not solely determined by the number of constraints and variables. This is observed in all four scenarios. When we vary $K$,
     \ac{MILP} requires a longer time to obtain a feasible solution and in many cases \emph{much longer time} to completely terminate the program. We hypothesize that it is affected by the large feasible solution space. Since we are maximizing the time robustness directly through the continuous timestamps, it is hard for the branch-and-bound methods to determine the optimal search path. 
  \end{enumerate}

\begin{table}    
\begin{center}
\begin{tabular}{ |c|c|c|c|c|c| } 
 \hline
 
\multicolumn{1}{|c|}{Methods} &\multicolumn{4}{|c|}{MILP Info} & \multicolumn{1}{|c|}{Time [s]} \\
 \hline 
\textit{PWL}/IB &\# Cstr & \# Bin & \# Cont & Obj ($L, \theta_+$) & GRB \\
\hline
\hline
\multicolumn{6}{|c|}{$\lgSpec_1$ - Multi-Blocks-1} \\
\hline
\textit{K = 23} &  22740 & 7724 & {895} & 44.7, 21.8 & {14*} 
\\
\textit{K = 25} & 25764 & 8810 & {975} & 44.7, 21.8 & {26*} - 165 \\
\textit{K = 30} & 34024 & 11805 & {1175} & 44.7, 21.8 & 165* - TO \\
T = 100 & 29214 & {7215} & 2959 & 44.7, 23 & 95* \\

\hline
\multicolumn{6}{|c|}{$\lgSpec_2$ - Multi-Blocks-2} \\
\hline
\textit{K = 25} & 49677 & {16636} & {1973} & 71.4, 17.7 & {38*} \\
\textit{K = 27} & 55371 & {18646} & {2187} & 70, 17.7 & {160*} - 195  \\
T = 300 & 192532 & 48275 & 16487 & 70, 17 & 630* - TO \\ 
\hline
\multicolumn{6}{|c|}{$\lgSpec_3$ - Door-Puzzle-1} \\
\hline
\textit{K = 23} & 18498 & 6394 & {648} & 75, 0.2 & 6* \\
\textit{K = 25} & 21209 & 7383 & {706} & 76, 9.7 & {27*} \\
\textit{K = 30} & 28739 & 10153 & {851} & 76, 9.7 & 149* - TO  \\
T = 100 & 21874 & {5537} & 2358 & 76, 10 & 78* - TO \\
\hline
\multicolumn{6}{|c|}{$\lgSpec_4$ - Door-Puzzle-2} \\
\hline
\textit{K = 35} & 39337 & {11855} & {2090} & 152, 10 & {251*} - 305 \\
\textit{K = 40} & 46577 & {14085} & {2395} & 143, 14.2 & {904*} - TO \\
T = 200 & 358929 & 111986 & 8566 & - & TO \\
\hline
\end{tabular}
\caption{Our method is denoted by the number of waypoints with the italic letter \textit{K}. Methods using discretized-signals are denoted by the planning horizon written in the plain text $T$. 
The \ac{MILP} complexity is reported through the number of linear constraints (\# Cstr), binary variables (\# Bin), and continuous variables (\# Cont), respectively. The objective $(L, \theta_+)$ indicates the optimized path length and the right time robustness. 
The last column depicts the optimization time of the Gurobi solver (in seconds). The number with the symbol $*$ denotes the time that the objective converges to the optimal solution. The number behind the hyphen (if exists) is the time that the solver needs to completely terminate the program. ``TO" is the time-out and is set to $1000$ seconds. 
}
\label{tab:bm}
\end{center}
\end{table}
To summarize, our method is \emph{sound} and the time-robust \ac{MILP} encoding with \ac{PWL} offers a complexity that is invariant to the planning horizon. This is especially advantageous for problems that can be represented by a few waypoints.
%
\section{Conclusion}  \label{sec:concl}
We synthesize time-robust trajectories that satisfy spatial-temporal specifications expressed in \ac{STL} using \ac{PWL} signals. To that end, we define a notion of temporal robustness and propose a sound quantitative \ac{STL} semantics for \ac{PWL} signals. Our encoding strategy leads to a \ac{MILP} encoding with fewer binary and continuous variables compared to the existing algorithms. This reduction in complexity of the constructed \ac{MILP} is more significant for \ac{STL} formulas with long time horizons. Our numerical experiments confirm the soundness of our semantics and our complexity analysis. 
Several limitations of this approach are the conservatism and the incompleteness of the semantics. Future works focus on analyzing the \ac{PWL} encoding of the atomic predicates to reduce the conservatism, as well as investigating the completeness property.



%

\bibliographystyle{IEEEtran}
\bibliography{refs.bib}

\begin{appendices}
\section{Soundness Proof of Time-robust STL for PWL Signal} \label{apx:sec_proof}
We prove Theorem \ref{theorem:soundness} 
induction hypothesis (IH) for each STL formula in the recursive syntax \eqref{eq:stl_syntax}. We prove the right-time robustness. The left one is derived similarly.
\begin{itemize}[leftmargin = *]
    \item Atomic predicate $\lgPred$ (similar for negation $\lgNeg \lgPred$)  
    From \eqref{eq:pwl_prop_tr_strict},
    ${\rtrobzj{i}^{\lgPred} > 0 \lgIff \sign(z_i^{\lgPred})\sum_{j = i + 1}^{i + k}(t_{j+1} - t_j) > 0.}$.
    Since ${(t_{j+1} - t_j) \geq 0 ~\forall j}$ and ${\sign(z_i^{\lgPred}) = \sign(z_{i^\prime}^{\lgPred}) \neq 0}$, we have
    ${\rtrobzj{i}^{{\lgPred}} > 0 \implies \sign(z^{\lgPred}) = 1 \lgIff  z_i^\pi = \lgTrue.}$    
    
    \item Conjunction: 
    \eqref{eq:log_time_rob_strict} ${\implies \trobzj{i}^{\lgSpec_1 \lgAnd \lgSpec_2} > 0 \implies (\trobzj{i}^{\lgSpec_1} > 0 \lgAnd \trobzj{i}^{\lgSpec_2} > 0)}$.
    From IH, we  have $z_i^{\lgSpec_1} = \lgTrue \lgAnd z_i^{\lgSpec_2} = \lgTrue$ ${\implies z_i^{\lgSpec_1 \lgAnd \lgSpec_2} = \lgTrue}.$

    \item Disjunction: \eqref{eq:log_time_rob_strict} ${ \implies \trobzj{i}^{\lgSpec_1 \lgOr \lgSpec_2} > 0 \implies (\trobzj{i}^{\lgSpec_1} > 0 \lgOr \trobzj{i}^{\lgSpec_2} > 0)}$
 IH implies $z_i^{\lgSpec_1} = \lgTrue \lgOr z_i^{\lgSpec_2} = \lgTrue$ ${\implies z_i^{\lgSpec_1 \lgOr \lgSpec_2} = \lgTrue}.$

    \item Until:
from \eqref{eq:until_time_rob},

${\trobzj{i}^{\phi_1\stlUntil{a,b}\phi_2} > 0}$
${\implies \exists j \in \set{J}^\intersect, \left( \trobzj{j}^{\phi_2} \infmum \binfmum_{l \in \set{L}_j^\intersect}\trobzj{l}^{\phi_1} \right) > 0}$  
${\implies \exists j \in \set{J}^\intersect, (\trobzj{j}^{\phi_2} > 0) \lgAnd \left(\forall l \in \set{L}_j^\intersect, \trobzj{l}^{\phi_1} > 0 \right) }$
${\implies \exists j, \bigg([t_j, t_{j+1}] \intersect [t_{i+1} + a, t_i +b] \neq \varnothing \lgAnd (z_j \lgSatisfy \lgSpec_2) \bigg) }$ 
${\lgAnd \bigg( \forall l \leq j, [t_l, t_{l+1}] \intersect [t_{i}, t_{i+1} +b] \neq \varnothing \lgAnd (z_l \lgSatisfy \lgSpec_1) \bigg)}$    
${\implies z_{i}^{\lgSpec_1\stlUntil{a,b}\lgSpec_2} = \lgTrue.}$   

Again, the second last implication is obtained from the induction rule with $\trobzj{l}^{\phi_1} > 0$ and $\trobzj{j}^{\phi_2} > 0$. A similar proof can be shown for "eventually" and "always" operators, and also for the ''violation" case, i.e., $\trobzj{i}^{\phi} < 0 \implies z_i^\phi = \lgFalse$.
 \end{itemize}

\end{appendices}

\end{document}